\documentclass{osa-article}

\journal{ome}

\articletype{Research Article}
\usepackage{lineno}

\begin{document}

\title{X-ray computed $\mu$-tomography for 
the characterization of optical fibers}

\author{Mario Ferraro,\authormark{1,2,\dag} Maria C. Crocco,\authormark{1,3,\dag,*}  Fabio Mangini,\authormark{2,\dag} Maxime Jonard,\authormark{4} Francesco Sangiovanni,\authormark{1} Mario Zitelli,\authormark{2} Raffaele Filosa,\authormark{1,3} Joseph J. Beltrano,\authormark{1,3} Antonio De Luca,\authormark{3} Riccardo C. Barberi,\authormark{1,3} Raffaele G. Agostino,\authormark{1,3} Vincent Couderc,\authormark{4} Stefan Wabnitz,\authormark{2,5} and Vincenzo Formoso\authormark{1,3}}

\address{\authormark{1}STAR Research Infrastructure, Università della Calabria, Via Tito Flavio, 87036 Rende (CS), Italy\\
\authormark{2}Department of Information Engineering, Electronics and Telecommunications, Sapienza University of Rome, Via Eudossiana 18, 00184 Rome, Italy\\
\authormark{3}Physics Department, Università della Calabria, Via Pietro Bucci, 87036 Rende (CS), Italy\\
\authormark{4}Universit\'e de Limoges, XLIM, UMR CNRS 7252, 123 Avenue A. Thomas, 87060 Limoges, France\\
\authormark{5}CNR-INO, Istituto Nazionale di Ottica, Via Campi Flegrei 34, 80078 Pozzuoli (NA), Italy.
\\

\authormark{\dag}These authors have contributed equally
}

\email{\authormark{*}mariacaterina.crocco@unical.it} 



\begin{abstract}

In spite of their ubiquitous applications, the characterization of glass fibers by means of all-optical techniques is still facing some limitations.
Recently, X-ray absorption has been proposed as a method for visualizing the inner structure of both standard and microstructure optical fibers. Here, we exploit X-ray absorption as nondestructive technique for the characterization of optical glass fibers. Starting from absorption contrast X-ray computed micro-tomography measurements, we obtain information about the spatial profile of the fiber refractive index at optical frequencies. We confirm the validity of our approach by comparing its results with complementary characterization techniques, based on electron spectroscopy or multiphoton microscopy. 
\end{abstract}

\section{Introduction}
Optical glass fibers are of widespread use in several technologies, ranging from telecommunications to lasers, endoscopy, and optical tweezers. Nowadays, optical fibers with a core radius of a few microns are mostly used. Thanks to their small core size, such fibers support the propagation a single guided mode at the wavelength of operation, hence they are named single mode fibers (SMFs) \cite{agrawal2000nonlinear}. Hence, laser light propagating in SMFs maintains a bell shaped beam at their output, which makes them suitable for many applications where high quality beams are needed. Nevertheless, SMFs present some drawbacks, e.g., they can only transport spatial information associated with one beam, and the pulse energy is limited by a relatively low threshold for nonlinear effects or even irreversible damage to take place. 

Therefore, in the pursuit for higher capacity optical network and higher power fiber lasers, multimode fibers (MMFs) have been captivating a renewed and growing interest over the last decade \cite{krupa2019multimode}. When compared with their single-mode counterparts, 
multiple transverse modes can 
simultaneously propagate in MMFs. This introduces an additional degree-of-freedom, which enables the multiplication of the information capacity of optical fiber networks.
As a consequence, MMFs have been proposed and demonstrated as a new enabling technology for telecommunications (through space- and mode- division multiplexing) \cite{ho2013linear}
, optical computing \cite{teugin2021scalable}, high power fiber lasers \cite{tegin2020single}, quantum information processing \cite{leedumrongwatthanakun2020programmable}, and optical imaging \cite{ploschner2015seeing}, e.g., for microscopy and endoscopy applications \cite{amitonova2018compressive, moussa2021spatiotemporal}. 


In this context, an important class of MMFs is provided by the so-called graded-index fibers (GIFs), whose refractive index has a parabolic profile. GIFs exhibit unique physical effects, such as spatial self-imaging \cite{karlsson1992dynamics, hansson2020nonlinear} and beam self-cleaning \cite{krupa2016observation,krupa2017spatial, liu2016kerr}. 
Parabolic ndex grading is produced by properly doping the fiber glass material: e.g., in silica fibers dopants such as Ge and F are typically used for increasing or decreasing the real part of the refractive index, respectively \cite{chen2009effects}. 

Besides standard optical fibers, i.e., fibers whose core and cladding are coaxial full cylinders, in the last decades several types of "specialty" optical fibers have been developed. These are made by shaping the inner structure of the fiber, giving rise to photonic crystal fibers (PCFs), multicore and multicladding fibers. Generally speaking, engineering the inner structure of optical fibers allows for extending the waveguiding mechanism of standard fibers beyond the principle of total internal reflection. PCFs have enabled the demonstration of efficient supercontinuum light sources \textcolor{red}{ \cite{dudley2006supercontinuum,genty2007fiber}}, based on the efficient engineering of their  chromatic dispersion profile \cite{saitoh2003chromatic}.

Regardless of their inner structure, the characterization of optical fibers mainly consists of the measurement of their refractive index profile. For such a purpose, several techniques have been developed,  which can be divided in two different categories: longitudinal and transverse techniques. \cite{stewart1982optical}.
To the first category belong non-optical techniques, such as chemical etching combined with atomic force microscopy \cite{zhong1994characterization}, as well as all-optical techniques, based on either the refraction \cite{young1981optical}, the reflection \cite{ikeda1975refractive, weng2015exploiting} or the interference of laser beams \cite{presby1976refractive}. Longitudinal techniques owe their name to the requirement of the on-axis coupling of optical beams to the fiber. Hence, fiber characterization by means of longitudinal techniques is based on the interaction of light with the fiber facet. Despite their efficiency, these techniques present several drawbacks, e.g., the need of a high-quality cleaved facet. Polishing the latter makes these methods inherently destructive, hence of difficult application to fiber-based devices. Moreover, longitudinal techniques provide information about the refractive index only in the vicinity of the fiber tip. Hence, these techniques do not allow for a mapping of the refractive index distribution throughout the entire fiber volume. 

On the other hand, transverse techniques involve optical beams which propagate through optical fibers perpendicularly to their axis. 
When compared to their longitudinal counterparts, transverse techniques have the main advantage that no constraints are posed to the quality of the fiber facet. This category includes phase-shifting interferometry \cite{sochacka1994optical}, quantitative-phase microscopy \cite{barty1998quantitative}, and Fourier-transform spectroscopy \cite{yablon2009multi}. Notably, particular interest has been attracted by micro-interferometric optical phase tomography \cite{bachim2005microinterferometric}. This combines transverse interferometry with optical tomography \cite{gorski2007tomographic}, thus providing a means for the full 3D determination of the fiber refractive index distribution \cite{fan2020reconstructing}.


In this framework, X-ray computed micro-tomography ($\mu$CT) has been recently proposed as a novel transverse technique for the investigation of optical fibers \cite{sandoghchi2014x}. At variance with optical tomography, which is based on the phase contrast that belongs to interference patterns, $\mu$CT relies on the absorption properties of the fiber at X-ray frequencies. Specifically, a $\mu$CT set-up may even use incoherent X-ray sources, by exploiting the absorption contrast between different materials. For this reason, $\mu$CT has conventionally found applications in fields where the absorption contrast is pretty high, such as biological images \cite{maugeri2018fractal} material science \cite{conte2019analysis}, geophysics, geology \cite{cnudde2013high,min12020177}, soil science \cite{taina2008application}, archaeology and cultural heritage \cite{stabile2021computational, lopez2021architectural}. When applied to fiber optics, absorption contrast $\mu$CT provides remarkable advantages when compared with optical tomography. For example, $\mu$CT can be carried out independently of the fiber bending, and even in the presence of a coating. Therefore, it is possible to carry out a $\mu$CT analysis without removing the fiber plastic jacket. However, it is worth to mention that, due to the low absorption coefficient of optical fibers at X-ray frequencies, accurate $\mu$CT measurements may require acquisition times as long as a few hours.

Preliminary studies of $\mu$CT of optical fibers have been reported in the literature. Specifically, $\mu$CT of PCFs and their preforms have been carried out by Sandoghchi et al. \cite{sandoghchi2014x}. Whereas, standard MMFs have been used by Levine and co-authors, in order to study the role of spatial averaging \cite{levine2019multi}, and to develop a $\mu$CT reconstruction algorithm including Fresnel diffraction \cite{levine2021x}. Nevertheless, to our knowledge a full investigation of MMFs by $\mu$CT has not been reported so far. 

In this work, we systematically apply $\mu$CT for analysis of several types of optical glass fibers, which are made by different manufacturers. Specifically, we propose a method that exploits $\mu$CT in order to determine the local values of different optical parameters, such as the core/cladding relative index and the grading factor of GIFs ($g$ parameter). The validity of our method is confirmed by means of comparisons with other complementary techniques, such as energy dispersed X-ray (EDX) spectroscopy, and multiphoton microscopy based on the luminescence of fiber defects. 



\section{Materials and methods}
\subsection{Optical fiber specimens}

In our experiments, we used 
spans of both microstructured and standard optical glass fibers. Specifically, our results involve the double-clad hypocycloid core-contour Kagome hollow-core PCF which is described in Ref. \cite{delahaye2018double}, as well as commercial GIFs and SIFs
, made by different manufacturers. 
A summary of our samples is reported in Table \ref{tab:fibre-list}. One may note that we did not use SMFs in our studies. We underline that this is because the size of their core is comparable with the spatial resolution of our experimental set-up. As a matter of fact, there are no intrinsic limitations to the possible use of $\mu$CT for the characterization of SMFs as well.

\begin{table}[!ht]
    \centering
\begin{tabular}{|c|c|c|c|c|c|}
\hline
    Sample & Type & Manufacturer & Diameters ($\mu$m) & $\mathit{NA}$ & Ref.\\
    \hline
    A & PCF & GLOphotonics & 44/143/187 & / & \cite{GLOphotonics}\\ 
    \hline
    B & SIF & Thorlabs & 50/125 & 0.22 & \cite{SIF50}\\ 
    \hline
    C & GIF & Alcatel & 50/125 & 0.2 & \cite{draka50}\\ 
    \hline
    D & GIF & Thorlabs & 50/125 & 0.2& \cite{GIF50e}\\ 
    \hline
    E & GIF & Thorlabs & 62.5/125 & 0.275 & \cite{GIF625}\\ 
    \hline
    F & GIF & Alcatel & 100/140 & / & \cite{draka100}\\ 
    \hline
    \end{tabular}
    \caption{List of optical fiber samples used in this work. The diameters refer to the core and cladding, respectively, while $\mathit{NA}$ stands for numerical aperture. All of the values are taken from the data-sheet provided by the manufacturer. In the case of sample A, the three values correspond to the inner and outer diameters of the smallest core, and the inner diameter of the widest cladding, respectively. 
    }
     \label{tab:fibre-list}
\end{table}


\subsection{Computed $\mu$-tomography set-up}

\begin{figure}[ht!]
\centering\includegraphics[width=8.7cm]{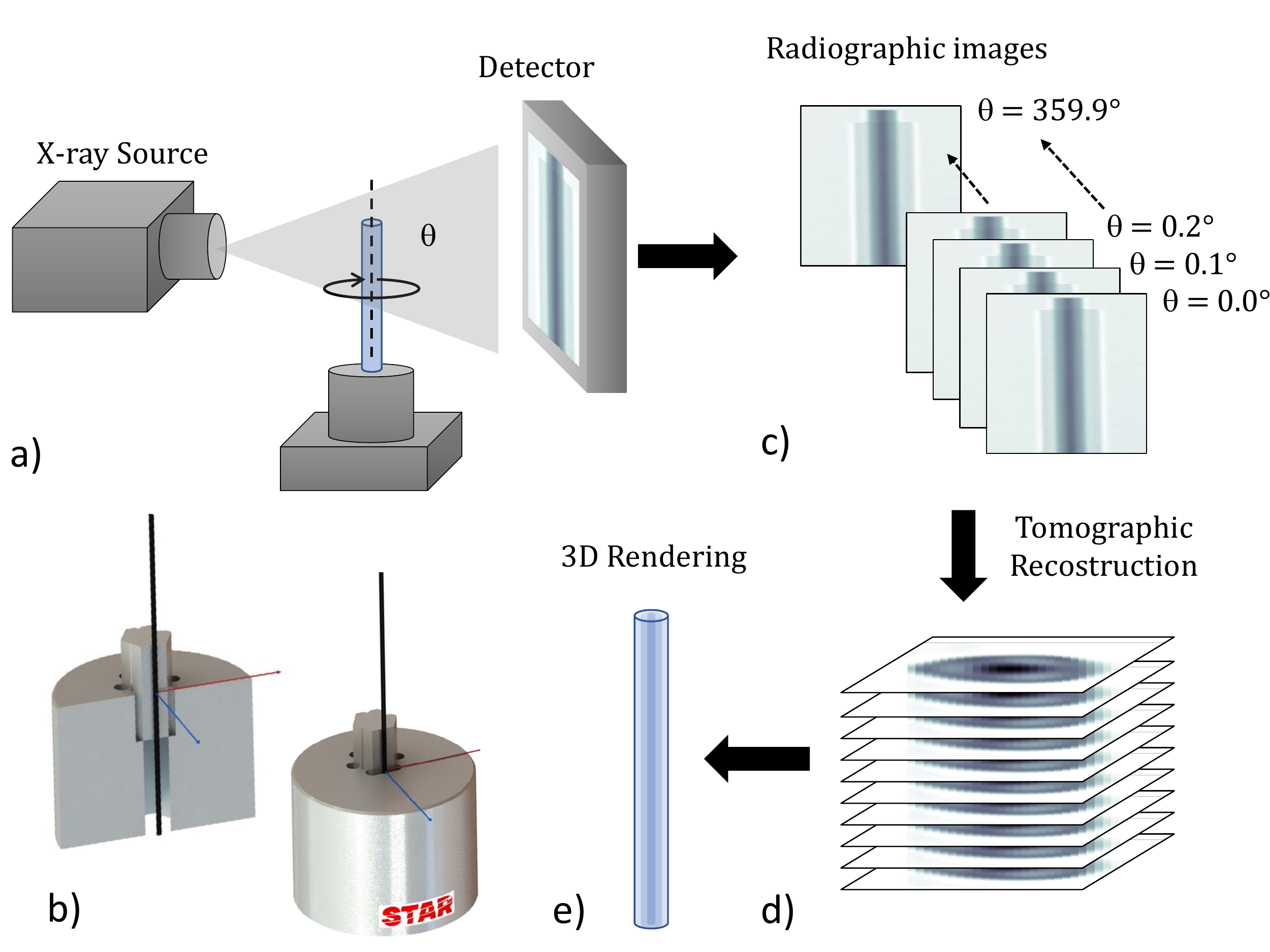}
\caption{a) Sketch of the $\mu$CT set-up. b) Computer-aided design images of an optical fiber sample, and its position with respect to the rotation axis of the holder. c) Set of radiographic images obtained for different rotation angles. d) Computed slices of the sample obtained via numerical routines based on the Feldkamp-Davis-Kress back-projection algorithm. e) Final result of $\mu$CT: reconstruction and 3D rendering of the sample.}
\label{fig:set-up}
\end{figure}

We use a microfocus source (Hamamatsu L12161-07), which emits a conical polychromatic X-ray beam with an aperture angle of $43 ^\circ$, and a focal spot of 5 $\mu$m. 
The source parameters are chosen so that 10 W of beam power, and a x-ray tube voltage of 60 kV. Radiographic images of the samples are captured by a flat panel detector (Hamamatsu C7942SK-05), as sketched in Fig.\ref{fig:set-up}a. The samples are located in a holder, which permits to overlap its own axis of rotation with the axis of symmetry of the fiber (see the computer-aided design in Fig.\ref{fig:set-up}b).

The measurements are carried out as follows. At first, a set of radiographic images is collected at different rotation angles $\theta$, as depicted in Fig.\ref{fig:set-up}c.
Next, by means of a numerical routine, the images are processed in order to obtain a stack of slices, which compose the tomographic reconstruction (see Fig.\ref{fig:set-up}d). Specifically, we relied on the Feldkamp-Davis-Kress back-projection algorithm, as described in Ref. \cite{Feldkamp:84}. Finally, by means of a rendering software, we obtain a 3D image of the sample (see Fig.\ref{fig:set-up}e). All of the images reported in this work are produced by means of both Fiji and Avizo software for image analysis and visualization \cite{schindelin2012fiji}.

It is worth to underline that the doses used in this work are sufficiently low to avoid any radiation induced attenuation of the optical transmission \cite{girard2004gamma, di2014x}. Indeed, \textcolor{black}{we verified that the exposition of the fibers to a dose of about 2000 Gy did not reduce their optical transmission. This result is coherent with previous observations reported in the literature, since } 2000 Gy is about one order of magnitude smaller than the value required for inducing modifications to the properties of our samples, because of X-ray absorption. This guarantees that $\mu$CT can be used as a non-destructive method for the characterization of optical fibers.

\subsection{Comparison Tomography / Radiography} \label{sec:radio-tomo}
One may wonder about the advantages of performing $\mu$CT, with respect to analysing a single radiographic image of optical fibers, as it is shown in Fig.\ref{fig:set-up}c. At first, since it involves acquisitions at different exposition angles, tomography permits to determine the presence of structural fiber asymmetries, as well as the occurrence of local modification to their refractive index, e.g. laser-induced damages \cite{ferraro2022multiphoton}. Notably, $\mu$CT allows for estimating the fiber non-circularity along the whole fiber axis, which may vary because of external mechanical stresses. Of course, it is impossible to extract this parameter by analysing a single radiographic image. Furthermore, by means of $\mu$CT one directly maps the absorption coefficient, without recurring to fitting processes. Indeed, when carrying out radiography, one gets a 2D profile of the transmitted intensity, which contains information about the absorption of a given sample, integrated all over its thickness. On the other hand, performing $\mu$CT permits to directly reconstruct a 3D map of the absorption properties (and consequently of the refractive index) of the sample, at the expense of increased measurement and computation time. As a matter of fact, $\mu$CT provides local information, while radiography only provides spatially averaged information. This is one of the main advantages of $\mu$CT, which permits to characterize, for example, the inner structure of an optical fiber which is coated by plastic jacket. We dedicate Appendix \ref{app-radio} to stress the differences between radiography and tomography. There, we show how and under which conditions one may extract information about the absorption coefficient, when starting from a radiographic image.

\subsection{Relationship between absorption coefficient and refractive index}

When considering both absorption and refraction properties of optical fibers, it proves convenient to introduce a single complex refractive index variable $\tilde{n} = n + i \kappa$. In fiber optics, the imaginary part $\kappa$ is often neglected, since the absorption of silica glass, which is related to $\kappa$, is rather low at optical frequencies \cite{agrawal2000nonlinear}. Hence, the refractive index is identified with its real part $n$ only. To the contrary, the role of $\kappa$ is crucial for $\mu$CT, since it is based on the absorption properties at X-ray frequencies. Specifically, $\mu$CT allows for a 3D mapping of the absorption coefficient ($\mu$), which is related to $\kappa$ by a simple linear proportionality relationship
\begin{equation}\label{eq:k-mu}
    \kappa \propto \mu. 
\end{equation}
It is worth to underline that this relationship only holds for monochromatic waves. 
However, at the X-ray frequencies used in this work, the chromatic dispersion of $\tilde{n}$ is negligible. Therefore, here we may suppose the direct proportionality between $\kappa$ and $\mu$, despite the polychromaticity of our source. Under this hypothesis, $\mu$CT measurements of $\mu$ can be identified with measurements of $\kappa$, within a multiplicative constant. In particular, as we will see in the next section, the parabolic profile of $\tilde{n}$ of GIFs allows for relating the 3D map of $\mu$, measured by $\mu$CT, with the 3D distribution of both $n$ and $\kappa$.
At this early stage, it is worth to underline that in this work we are relating the values of $\kappa$ and $n$ which belong to distant spectral ranges: X-rays and infrared radiation, respectively. Owing to such a big spectral gap, $n$ cannot be computed by starting from the measurements of $\kappa$ via Kramers-Kronig formulas, as it is commonly done for calculating $n$ in the vicinity of X-ray frequencies.

\section{Results}\label{sec:results}
\subsection{Computed $\mu$-tomography of different fiber types}
As a first step, we performed a tomographic reconstruction of different types of optical glass fibers. In Fig.\ref{fig:tomography}a, b, and c, we show the case of samples A, B, and C, respectively, whose data for the 3D reconstruction were collected within the same experimental conditions (see also visualization 1, 2, and 3 in the Supplementary Materials). Specifically, all of these specimens were exposed to X-radiation for a total time of 20 h, while rotating with steps of $0.1^\circ$, from $0^\circ$ to $360^\circ$. In each step, the radiographic image was averaged over 3 acquisitions. A magnification of 11 was obtained, by placing the sample holder at a distance of 10 cm and 100 cm from the source and the detector, respectively. Such distances provide the optimal conditions for $\mu$CT measurements, giving the maximum resolution achievable by our set-up, i.e., of about 5 $\mu$m.

By visually comparing the three samples shown in Fig.\ref{fig:tomography}, one may immediately appreciate that $\mu$CT is able of fully capturing their structural features. Indeed, in Fig.\ref{fig:tomography}a the double cladding structure of the hollow core PCF is clearly appreciable. One can recognize both the outer and inner cladding, as well as the radial structure which sustains the latter. Whereas for standard MMFs, besides the geometry of the sample, $\mu$CT is capable of capturing information about the refractive index difference between core and cladding. By comparing Fig.\ref{fig:tomography}b and c, one may easily visually appreciate the difference between SIFs and GIFs. The inner part of the former is rather homogeneous, and a chromatic jump occurs in the periphery. This is the hallmark of the refractive index stepwise variation, when passing from the core to the cladding. On the other hand, the GIF sample is characterized by a progressive color variation when moving from the fiber axis towards its edges: this can be ascribed to the parabolic grading of its refractive index.

\subsubsection{Fiber non-circularity}
\begin{figure*}[!ht]
\centering\includegraphics[scale=0.5]{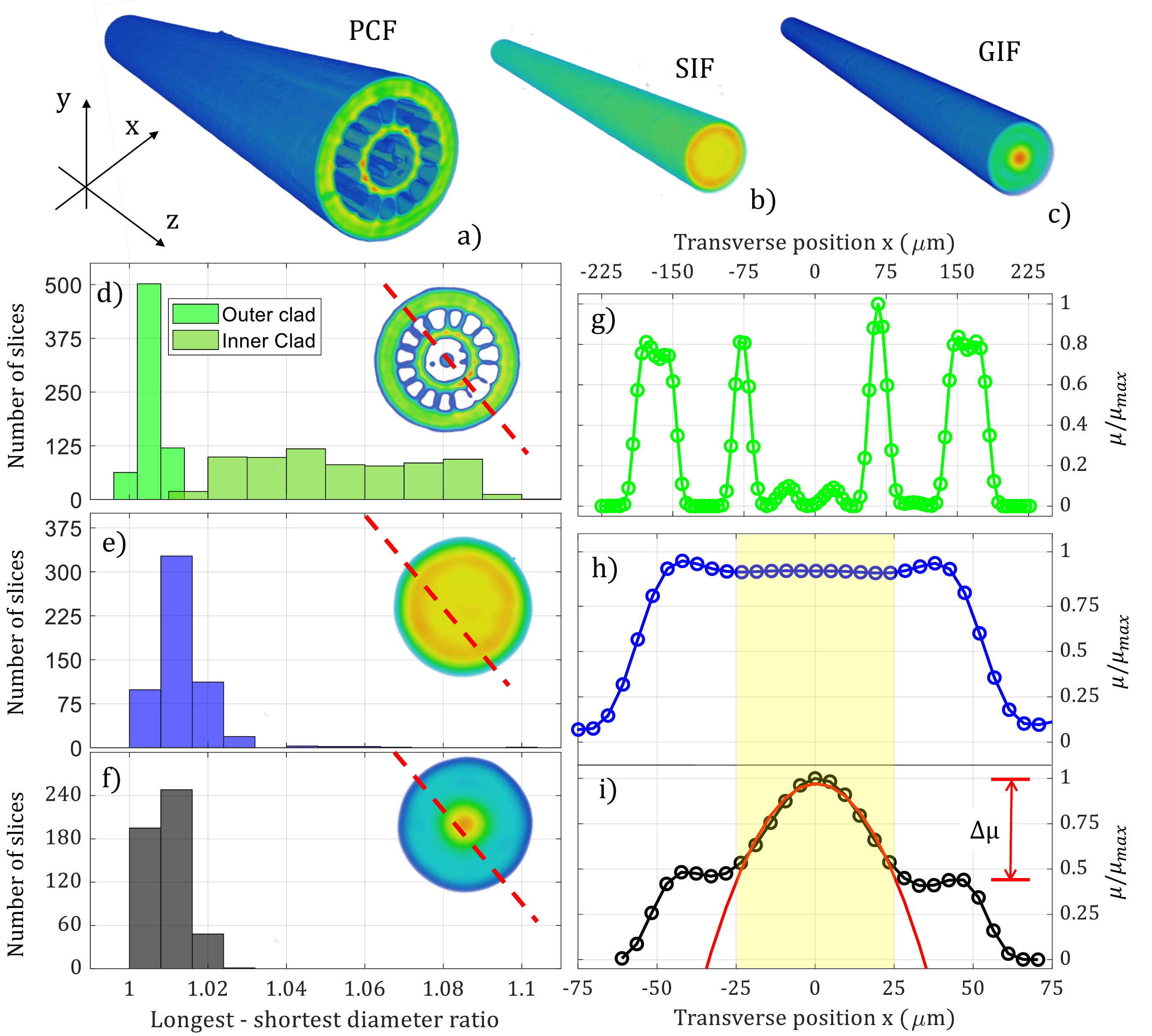}
\caption{(a-c) Tomographic 3D reconstruction of samples A, B, and C in the reference system shown in the bottom. (d-f) Statistical occurrence of the ratio between the longest and the shortest diameter
. The inset shows a single 2D slice of the 3D reconstruction in (a-c). (g-i) Corresponding profile along the dashed red lines in (d-f). Note that the images in (a-c) are normalized to the same values, i.e., they share the same color code. Whereas the profiles in (g-i) are normalized to their respective own maximum. The yellow area highlights the core size, while the red solid curve in (i) is a parabolic fit to the experimental data.
}
\label{fig:tomography}
\end{figure*}

Additional differences among the considered fiber samples can be appreciated by looking at their single reconstructed slice, as shown in the inset of Fig.\ref{fig:tomography}d-f. It turns out that SIF and GIF samples are rather circular, whereas the PCF shows a slight asymmetry of its diameter. This is highlighted in the main plot of Fig.\ref{fig:tomography}d-f, where we show the statistical distribution of the ratio between the longest and the shortest diameter, for each reconstructed slice. The latter are calculated by binarizing all of the slices, i.e., by associating either zero or one to pixels whose intensity is below or above a given threshold, respectively. Of course, being the PCF a double-clad fiber, in Fig.\ref{fig:tomography}d we report statistics for both the inner as well as the outer cladding. As it can be seen, there is a rather clear difference among the samples. The diameter ratio distribution associated to a SIF ora a GIF peaks at 1.01. This is coherent with the nominal value of fiber non-circularity of 1\%, as provided by the fiber manufacturer. Similar values are obtained for the outer cladding of the PCF. Whereas we obtained a rather different distribution of the diameter ratio for the inner clad of the PCF. As it can be seen in Fig.\ref{fig:tomography}d, the diameter ratio distribution peaks around 1.05. This permits to estimate that the non-circularity of the inner cladding is about five times higher than that of the outer cladding. Furthermore, it is worth to underline that the distribution of the diameter ratio for the inner cladding is larger than all for all other cases which are reported in Fig.\ref{fig:tomography}d-f. This indicates that the inner cladding is rather patchy along $z$, probably due to the radial structure which holds it. Whereas the outer diameter of the PCF, the SIF, and the GIF turns out to be quite uniform in all of the reconstructed slides. 

\subsubsection{Refractive index mapping}
Next, we extract the profile of the reconstructed $\mu$CT image, i.e., the variation of the absorption parameter $\mu$, along the fiber diameter (see the red dashed line in Fig.\ref{fig:tomography}d-f). The profiles of $\mu$, normalized to their peak values $\mu_{max}$, are reported in Fig.\ref{fig:tomography}g-i for samples A, B, and C, respectively. As it can be seen from Fig.\ref{fig:tomography}g, the inner Kagome structure of sample A is not clearly distinguishable. This is because the latter is rather thin: its typical size is of about a few microns, which is below the resolution achieved by our $\mu$CT set-up.  
Moreover, one may notice that $\mu$CT measurements are not capable of fully resolving the sharp difference between the cladding and the air absorption coefficients, which is expected to ideally produce a step in the blue curve in Fig.\ref{fig:tomography}h. To the contrary, the latter shows a relatively slow variation. This indicates that absorption is not the only physical mechanism leading to a variation of the transmitted intensity profile. Indeed, other effects such as refraction and diffraction may play a non-negligible role.
Nevertheless, as long as we analyse data collected far from interfaces, e.g., when limiting the  analysis to the fiber core (which is highlighted by a yellowish rectangle in Fig.\ref{fig:tomography}h,i), one can reasonably assume that absorption is the dominating mechanism for producing radiography and, consequently, to $\mu$CT images.
We underline that the scale of the horizontal axis in Fig. \ref{fig:tomography}g differs from that of Fig. \ref{fig:tomography}h and i. As a matter of fact, the spatial resolution of $\mu$CT remains the same for both PCFs and for standard MMFs.

\subsubsection{The role of doping}
One may notice that the blue curve in Fig.\ref{fig:tomography}h is flat inside the core. Whereas, when moving from the fiber axis to the edges, the same curve slightly grows, before dropping to zero at the cladding/air interface. This slight increase is ascribable to the presence of F doping inside the cladding. This enhances the absorption at X-ray frequencies (i.e., $\kappa$) and, at the same time, reduces the value of $n$, thus ensuring optical beam waveguiding \cite{chen2009effects}.
Finally, in sample C, the $\mu$CT trace is higher in the core than in the cladding. This is due to the presence of Ge doping, which provides the index grading of the core, and thus increases both $n$ and $\kappa$.
The doping of optical fibers is crucial for their characterization by means of $\mu$CT, since it is responsible for the absorption contrast which allows for distinguishing between different materials, e.g., the core from the cladding of SIFs. In the specific case of GIFs, whose core complex refractive index ($\tilde{n}_{core}$) can be expressed as
\begin{equation} \label{eq:n-gif}
   \tilde{n}_{core}(x,y) = \tilde{n}_0 \Big[1-g(x^2+y^2)\Big],
\end{equation}
$\mu$CT turns out to be able to detect the refractive index grading, as it can be seen in Fig.\ref{fig:tomography}i. Specifically, inside the fiber core, the $\mu$CT trace exhibits a bell-shaped profile, which is highlighted by a parabolic fit (red solid line in Fig.\ref{fig:tomography}i).

\subsection{Validity of the method}
In writing Eq.(\ref{eq:n-gif}), we assume that both the imaginary part of the refractive index at X-ray frequencies and the real part of the refractive index at optical frequencies share the same dependence on the transverse position ($x,y$). This hypothesis is at the basis of our method, since it permits to relate the spatial profile of $n$ and $\kappa$ to that of $\mu$ through Eq.(\ref{eq:k-mu}). Because of the crucial relevance of this hypothesis, we have verified his validity. In order to do so, as we shall see in the following, we compared the $\mu$CT analysis with EDX measurements, as well as with results extracted from the luminescence of fiber defects.

\subsubsection{Comparison of tomography with EDX measurements}

\begin{figure}[ht!]
\centering\includegraphics[width=8.7cm]{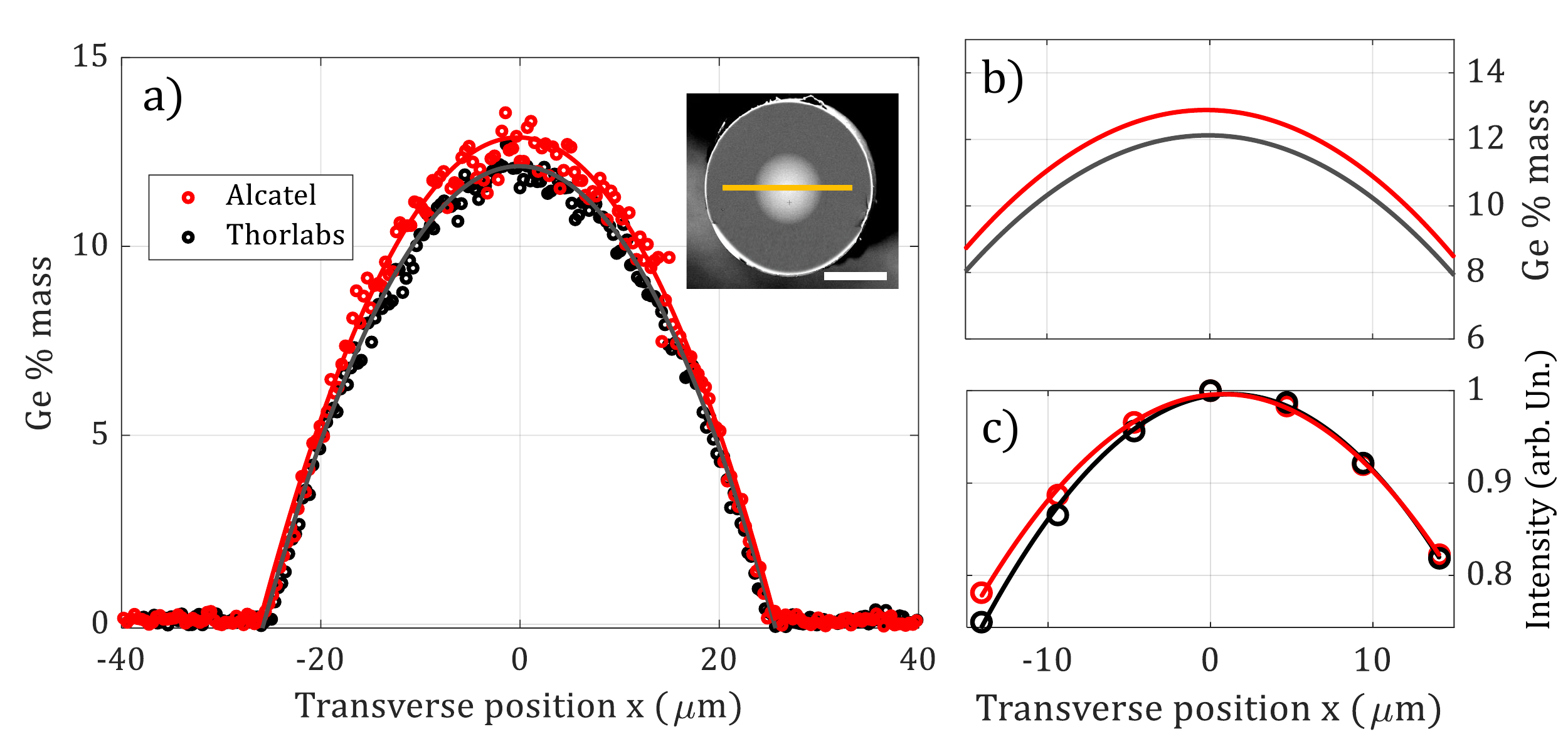}
\caption{Comparison between $\mu$CT and EDX measurements of samples C and D, marked as "Alcatel" and "Thorlabs", respectively. a) Ge concentration (\% in mass) along the yellow line traced on the SEM image in the inset. The white bar is 50 $\mu$m long. b) Zoom-in of the parabolic fit curves in a). c) $\mu$CT experimental data (void circles) and corresponding parabolic fit of the same samples in a).
}
\label{fig:EDX}
\end{figure}

EDX spectroscopy allows for tracing the characteristic X-ray emission of the dopant \cite{shindo2002energy}. 
Such emission is triggered by the ejection of deep inner shell electrons, caused by the incidence of the high-energy electron beam of a scanning electron microscope (SEM). 
In this contest, EDX spectroscopy permits to verify that the spatial profile of $\kappa_{core}$ follows that of the doping concentration at the fiber facet. Here, we analyze samples C and D, which are both GIFs with the same core and cladding sizes, and the same value of $\mathit{NA}$ (cfr. Table \ref{tab:fibre-list}). Moreover, their index grading is realized by means of the same dopant, i.e., Germanium. A typical SEM image of the optical fiber facet is shown in the inset of Fig.\ref{fig:EDX}a. We used a Zeiss crossbeam 550 SEM, equipped by an EDX spectrometer. The electron beam has 15 keV of energy, which allows for distinguishing the characteristic radiation emitted by Ge atoms, whose $L_\alpha$ emission peaks at 1.188 keV. By tracing the intensity of the latter along the yellow line in the inset of Fig.\ref{fig:EDX}a, we obtained the doping concentration (\% in mass) shown in the main plot of the same figure. One may notice that both samples have a maximum Ge concentration at the core center, which progressively reduces, until it disappears when approaching the core/cladding interface. Experimental data for both samples are well fit by a parabola (see the solid lines in Fig.\ref{fig:EDX}a). 
We highlight that both fitting curves cross the $x$-axis at the same points, thus confirming that the two samples share the same core size. On the other hand, the fitting curves have different peak values (as highlighted in Fig.\ref{fig:EDX}b). 

As a matter of fact, the profile extracted from the EDX analysis, i.e., the Ge concentration in Fig.\ref{fig:EDX}a and b, has the same (parabolic) profile at that extracted from $\mu$CT images, which is shown in Fig.\ref{fig:EDX}c. Specifically, the ratio between the convexity parameter of the parabolic fit, which is proportional to the ratio between the grading factor $g$ of the two fibers, turns out to be fully consistent when both methods are used. Indeed, we estimate a discrepancy of less than 5\% between the ratio between the values of $g$ which are calculated from either EDX or $\mu$CT measurements. 

The consistency between these two methods highlights a main benefits of $\mu$CT with respect to all-optical techniques for optical fiber characterization. Similarly to EDX, $\mu$CT permits to distinguish small differences between optical fibers (made by different manufacturers) which have the same nominal optical properties. 

\subsubsection{Estimation of the core/cladding relative index}
\begin{figure}[ht!]
\centering\includegraphics[width=8.7cm]{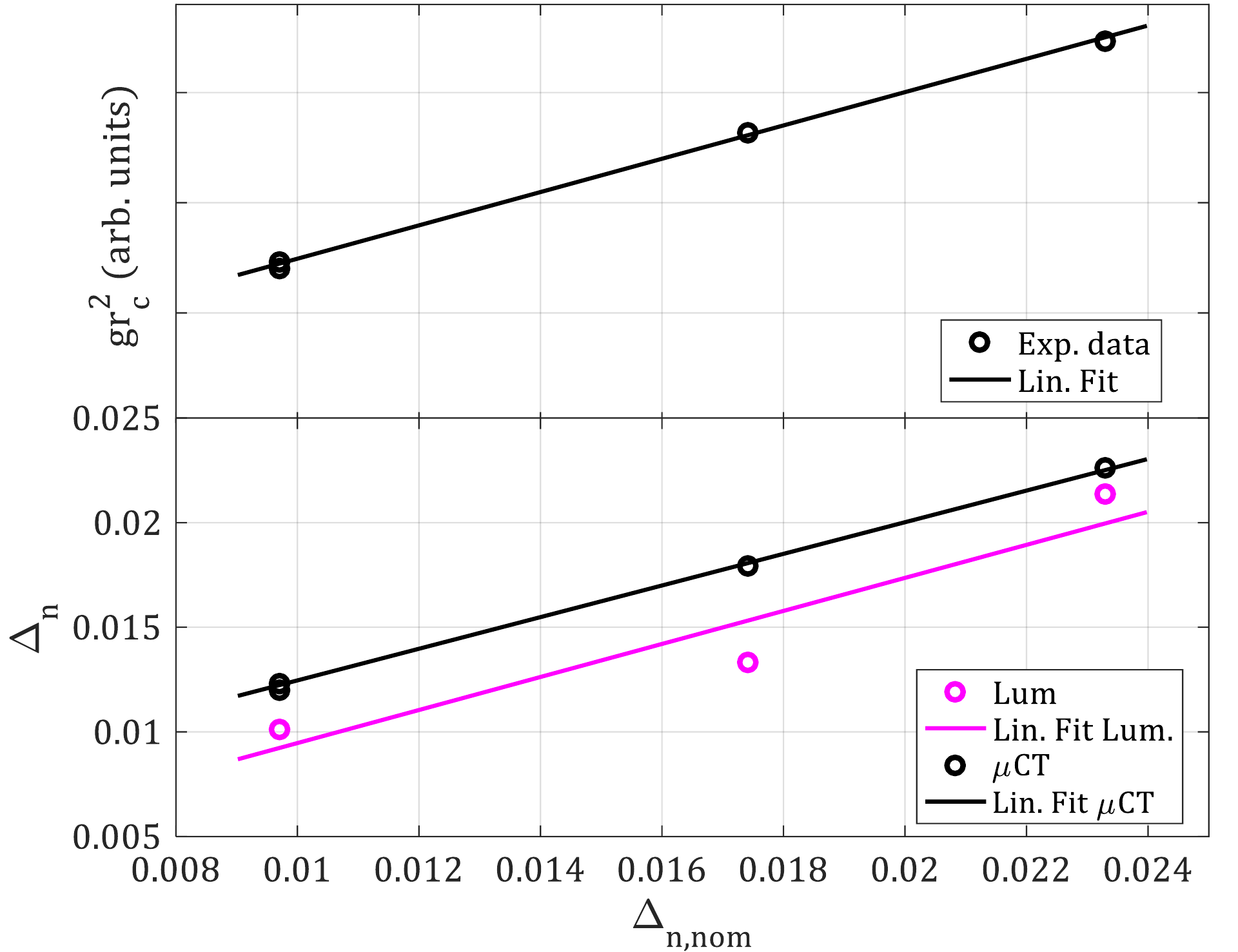}
\caption{(a) Experimental values of $g r_c^2$, calculated by fitting the $\mu$CT profile vs. $\Delta_{n,nom}$ for samples C, D, E, and F. (b) Values of $\Delta_n$ calculated from either the measurement of $\Delta\mu$ in the $\mu$CT intensity profile (black circles), or from the analysis of the fiber luminescence (purple circles), vs. $\Delta_{n,nom}$. Solid lines are linear fits of the experimental data.
}
\label{fig:lum}
\end{figure}

Next we verified another consequence of the assumption at the basis of our method, i.e., that measuring the imaginary part of the core/cladding refractive index difference $\Delta_\kappa$ at X-ray frequencies allows for estimating its real part $\Delta_n$ at optical frequencies. This can be demonstrated by imposing the condition $\tilde{n}_{core} = \tilde{n}_{clad}$ in Eq.(\ref{eq:n-gif}) whenever $x^2+y^2=r_c^2$, where $r_c$ is the core radius, which yields for GIFs
\begin{equation}
    \tilde{n}_0 g r_c^2 = \tilde{n}_0-\tilde{n}_{clad} \equiv \Delta_n + i \Delta_\kappa.
\end{equation}
By separating real and imaginary parts of the left and right-hand sides of Eq.(3), immediately yields that both $\Delta_n$ and $\Delta_\kappa$ are proportional to the \emph{real} quantity $g r_c^2$, since
\begin{equation}
    g r_c^2 = \frac{\Delta_n}{n_0} = \frac{\Delta_\kappa}{\kappa_0},
\end{equation}
where $n_0$ and $\kappa_0$ are the real and imaginary parts of the refractive index at the center of the core, respectively. Note that $g r_c^2$ can be estimated via $\mu$CT, by following the same procedure described in the previous section. With this method, we calculate the value of the parameter $g$ by fitting the $\mu$CT profile along one diameter of the fiber slice. Specifically, we considered all of the four GIF samples in our palette, i.e., samples C, D, E, and F. In Fig.\ref{fig:lum}a, we plot as void circles the values of $g r_c^2$ as a function of the nominal value of $\Delta_n$, i.e., that obtained starting from the information provided by the fiber manufacturer, which is dubbed $\Delta_{n,nom}$. The values of the latter are calculated from the $\mathit{NA}$ values in Table \ref{tab:fibre-list}. The calculation procedure is described in the Supplementary Materials, whereas a list of the values of $\Delta_{n,nom}$ is reported in Table \ref{tab:delta}. As it can be seen in Fig.\ref{fig:lum}a, the experimental data are well interpolated by a straight line (the linear correlation coefficient is $r^2 = 0.9958$). 

Finally, we demonstrate that a relationship of linear proportionality holds between $\Delta_n$ at optical frequencies and $\Delta_\kappa$ at X-ray frequencies (as it is measured by $\mu$CT). Indeed, according to Eq.(\ref{eq:k-mu}), by measuring the difference of the absorption coefficient between core and cladding ($\Delta\mu$), we may determine $\Delta_\kappa$, within a multiplicative constant. Therefore, in order to achieve our goal, it is sufficient to prove a direct proportionality between $\Delta_n$ and $\Delta_\mu$, which can be estimated as indicated in Fig.\ref{fig:tomography}i. This is shown in Fig.\ref{fig:lum}b, where the value of $\Delta_n$, obtained from $\Delta\mu$ measured by $\mu$CT multiplied by a constant vs. $\Delta_{n,nom}$, are shown as black void circles. As it can be seen, the fit of experimental points by a linear relationship is excellent (we found a linear correlation coefficient of 0.9998). Note that the value of the multiplicative constant that we used for calculating $\Delta_n$ starting from the measurement of $\Delta\mu$ was chosen so that $\Delta_n$ of sample C coincides with its nominal value. In this way, sample C acts as a reference sample and, by means of $\mu$CT, we could also determine the values of $\Delta_n$ for samples D,E and F, which are reported in Table \ref{tab:delta}.

It is worth underlining that, similarly with the estimation of $g$ in the previous section, $\mu$CT also allows for distinguishing the value of the core/cladding relative index of fibers with the same nominal parameters, notably samples C and D (cfr. the two couples of black circles with the same value of $\Delta_{n,nom}$ in Fig. \ref{fig:lum}b.

\subsubsection{Luminescence of fiber defects}

Finally, we may compare the characterization of optical fibers that we made by means of $\mu$CT, with the results extracted from a study of the luminescence of fiber defects. The latter is a technique for indirectly gathering information about the propagation of laser beams inside the fiber, as well as about the fiber characteristics \cite{girard2005luminescence, girard2019overview}. For example, fiber luminescences has been recently exploited to track the propagation path of helical beams \cite{mangini2022helical, mangini2020spiral}, as well as for estimating the frequency cutoff of SMFs \cite{mangini2021experimental}. 
Information about the physical phenomena behind the excitation of fiber luminescence, and a scheme of the experimental set-up for its characterization, are reported in the Appendix \ref{app-lum}. The values of $\Delta_n$ which are calculated from the analysis of the luminescence are reported in Table \ref{tab:delta}: their dependence on $\Delta_{nom}$ is shown in Fig.\ref{fig:lum} (bottom panel) by means of purple void circles. Here, we compare samples C, D, E, and F: these are all GIFs, and have different core sizes. For $\mu$CT measurements, we found that a linear interpolation fits well the experimental data (see the a purple solid line in the bottom panel of Fig.\ref{fig:lum}). However, the linear correlation coefficient associated with the fit of the luminescence data turns out to be 0.9231, which is a relatively small value when compared with the result of the $\mu$CT analysis. 
Besides, we underline that the use of $\mu$CT has significant advantages with respect to the luminescence-based analysis. The latter, in fact, may only be exploited close to the fiber tip, typically over distances of a few millimeters. After such a distance, the luminescence signal is damped by both linear and nonlinear losses \cite{ferraro2021femtosecond}. On the other hand, $\mu$CT analysis can be performed at any distance from the fiber tip, and even in the presence of a plastic jacket, a situation that would hinder any luminescence. \textcolor{black}{In this regard, it is worth noticing that one of the main advantages of $\mu$CT is that it is possible to measure long pieces of fiber at once, e.g., by coiling them instead of using the straight fiber configuration of Fig. \ref{fig:set-up}.}


\begin{table}[!ht]
    \centering
\begin{tabular}{|c|c|c|c|}
\hline
\multicolumn{4}{|c|}{$\Delta_n$} \\
\hline
    Sample & Nominal & $\mu$CT & Luminescence \\
    \hline
    D & 0.0097 & 0.0123 & 0.0101 \\
    \hline
    E & 0.0174 & 0.0179 & 0.0133 \\
    \hline
    F & 0.0233 & 0.0226 & 0.0214 \\
    \hline
    \end{tabular}
    \caption{Comparison of the core/cladding refractive index $\Delta_n$ measured by different techniques. 
    }
     \label{tab:delta}
\end{table}

\section{Conclusions}

In conclusion, we in this work we have shown that $\mu$CT is a technology which is capable of capturing the main characteristics of optical fibers. By exploiting the absorption contrast at X-ray frequencies of the different materials that constitute the fiber core and cladding, one is able of visualizing their full 3D inner features. This is particularly effective for PCFs, whose photonic crystal structure enhances the absorption contrast.
As far as standard optical fibers are concerned, in this work we only investigated the properties of MMFs. Nevertheless, all of the results reported herein can also be applied to SMFs, as soon as a $\mu$CT set-up with sufficiently high spatial resolution is employed.

We found that $\mu$CT allows for clearly visualizing the differences between SIFs and GIFs. The latter were used for verifying the reliability of the $\mu$CT analysis. Specifically, we found an excellent consistency between $\mu$CT and EDX measurements of the doping concentration in the fiber core (which is responsible for index grading), as well as with characterization results based on the luminescence of fiber defects. We have shown that $\mu$CT can be used for determining both the core/cladding relative refractive index values, and the grading factor of GIFs. 

As a matter of fact, using the $\mu$CT technique has several advantages, when compared with other techniques for optical glass fiber characterization. Notably, $\mu$CT allows for simultaneously determining both geometrical (e.g., fiber non-circularity as well as core, cladding, and coating sizes) and optical properties of the fibers. Remarkably, the $\mu$CT analysis can be carried out on long spans of optical fibers, thus providing information on the entire sample volume, which is not the case when EDX or luminescence are amployed. Furthermore, $\mu$CT does not require the sample to be put under ultra-high vacuum (as for EDX spectroscopy), and it works regardless of the presence of fiber bending or a plastic coating.

All of these considerations confirm that $\mu$CT is a valuable tool for optical fiber characterization. From a fundamental point of view, our results pave the way for future investigations using a phase-contrast $\mu$CT set-up, which can provide higher quality images when compared with those obtained by absorption contrast \cite{mayo2003x}. Of course, this would require the use of coherent X-ray sources, such as synchrotrons\textcolor{black}{, which allow for reaching resolution below one micron \cite{withers2021x}.}

\section*{Funding}
Ministero dell’Istruzione, dell’Università e della Ricerca (PIR01-00008, R18SPB8227); European Research Council (740355); Agence Nationale de la Recherche (ANR-18-CE080016-01, ANR-10-LABX-0074-01); Sapienza University of Rome Progetti di Avvio alla Ricerca (AR22117A7B01A2EB, AR22117A8AFEF609).

\section*{Acknowledgements}
We acknowledge the support of CILAS Company (ArianeGroup, X-LAS laboratory) and “Région Nouvelle Aquitaine” (F2MH and SI2P).

\section*{Disclosures}
The authors declare no conflict of interest.
\section*{Data availability}
Data underlying the results presented in this Article are not publicly available at this time but may be obtained from the authors upon reasonable request.


\bibliography{biblio}
\newpage
\noindent {\Huge \textbf{Supplementary Materials}}
\appendix

\section*{Analysis of radiographic images}\label{app-radio}

\begin{figure}[ht!]
\centering\includegraphics[width=8.7cm]{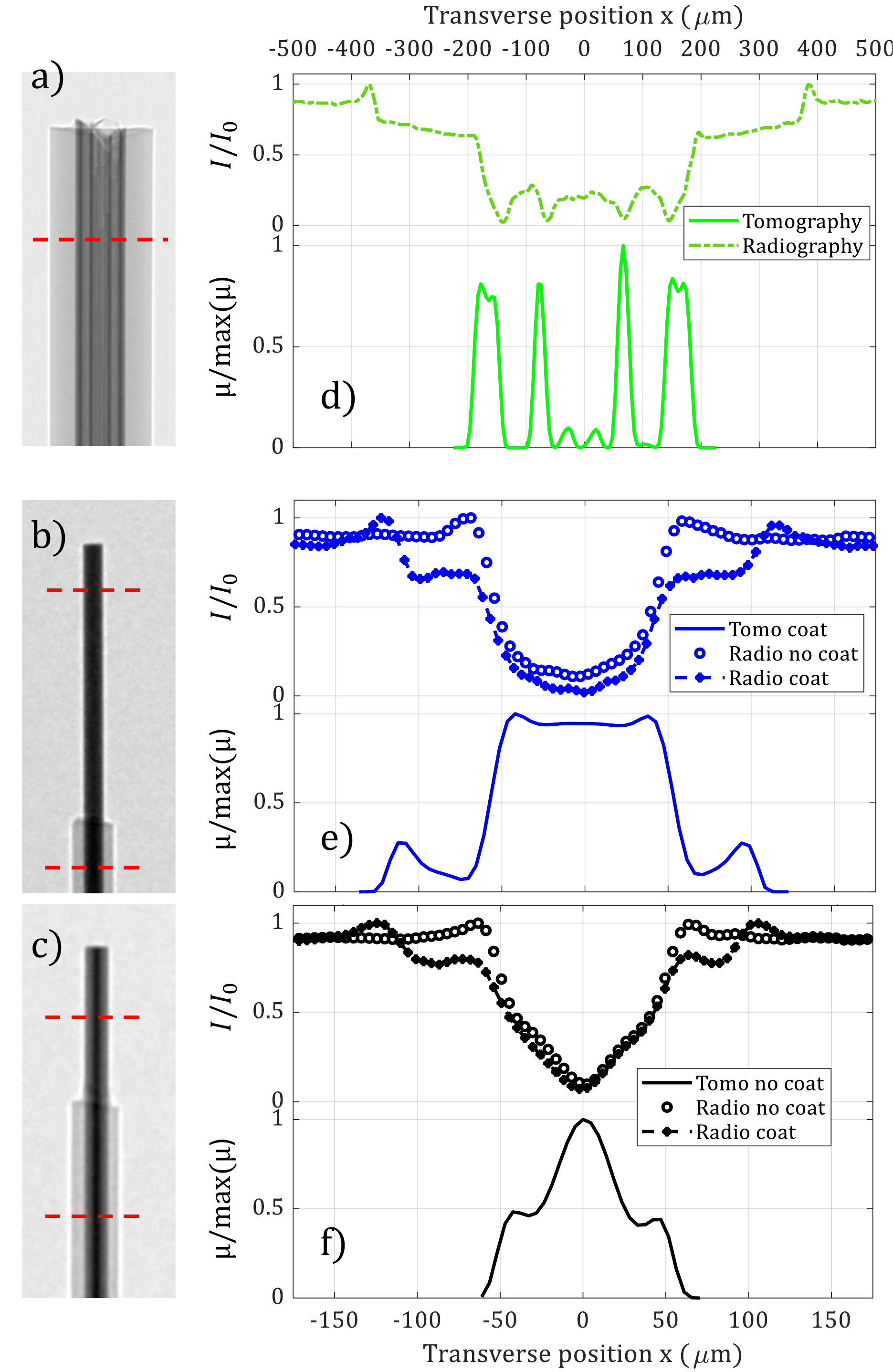}
\caption{(a-c) Radiography of sample A (a), B (b) and C (c). All of the dashed red lines are 500 $\mu$m long. (d-f) Comparison between tomography profile (solid lines) with the intensity profile extracted from the red dashed lines in (a-c).
}
\label{fig:radiography}
\end{figure}

In this section, we present the analysis of radiographic images of SIF, GIF, and PCF. Specifically, we refer to the same samples reported in Fig.\ref{fig:tomography}, i.e., samples A, B, and C, whose radiographic images are reported in Fig.\ref{fig:radiography}a, b, and c, respectively. All of these images were obtained by averaging over 100 acquisitions, corresponding to a total exposure time of 700 s.

As a first step, we compare the profile extracted from the tomographic images which are reported as solid lines in Fig.\ref{fig:radiography}d, e, and f with that extracted from radiographic images. As mentioned in the main text, these two types of images carry different information. A tomographic profile provides the absorption coefficient $\mu$, whereas radiographic images show the profile of the transmitted intensity. Therefore, when normalized to their respective maximum values, these profiles look complementary, i.e., when one reaches its maximum, the other has a minimum, and vice-versa. This can be clearly seen in Fig.\ref{fig:radiography}d, e, and f, where the radiographic trace are extracted along the red dashed lines in Fig.\ref{fig:radiography}a, b, and c. In particular, for SIFs and GIFs, we extract two different profiles, which either include or exclude the presence of plastic coating. This permits to see that the two profiles qualitatively differ in the cladding and core regions. Still, little differences are visible in the case of a SIF sample (see Fig.\ref{fig:radiography}e): these are ascribable to a normalization issue. This confirms that X-ray absorption is a valuable tool for investigating the inner properties of optical fibers, regardless of the presence of a coating. 




\begin{figure}[ht!]
\centering\includegraphics[width=8.7cm]{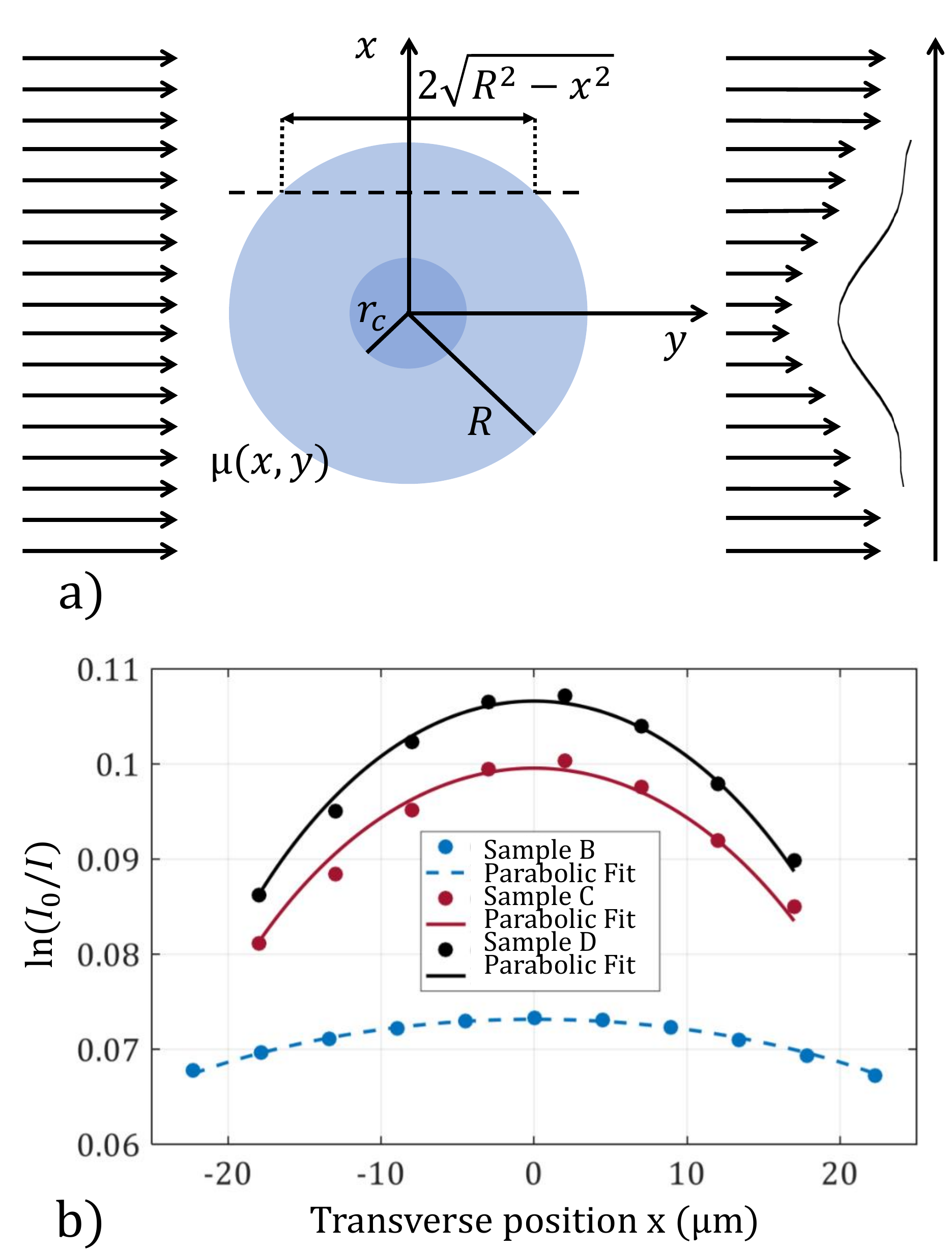}
\caption{(a) Scheme of the geometry of X-ray absorption by an optical fiber. (b) Fit of the radiographic image trace of samples B, C, and D by means of Eq.(\ref{eq:fit}). The fit parameters are $\mu_0 = 6.08 \cdot 10^{-4}$ $\mu$m$^{-1}$, $\mu_R = 5.7 \cdot 10^{-4}$  $\mu$m$^{-1}$ for sample B, $\mu_0 = 1.3 \cdot 10^{-4}$ $\mu$m$^{-1}$, $\mu_R = 4.1 \cdot 10^{-4}$ $\mu$m$^{-1}$, $g_\mu = 2.63 \cdot 10^{-3}$ $\mu$m$^{-2}$ for sample C and $\mu_0 = 8.5 \cdot 10^{-5}$ $\mu$m$^{-1}$, $\mu_R = 4.05 \cdot 10^{-4}$ $\mu$m$^{-1}$ and $g_\mu = 3.7 \cdot 10^{-3}$ $\mu$m$^{-2}$ for sample D, respectively.
}
\label{fig:theo}
\end{figure}

We present now a model which is able to fit the experimental data acquired by a single radiography. Let us consider a standard optical fiber without any coating, whose section can be sketched by two concentric circles with radius $r_c$ and $R$, respectively (see Fig.\ref{fig:theo}a). The inner circle represents the core, whereas the outer circle represents the cladding. By using the coordinate system in Fig.\ref{fig:theo}a, we can write the absorption coefficient of the optical fiber as:
\begin{equation}\label{eq:mu}
    \mu(x,y) = \left\{ \begin{array}{lcr}
        0 & \textrm{if } & \sqrt{x^2+y^2}>R \\
        \mu_R & \textrm{if } & r_c<\sqrt{x^2+y^2}<R\\
        \mu_c(x,y) & \textrm{if } & \sqrt{x^2+y^2}<r_c
    \end{array}
    \right. ,
\end{equation}
where $\mu_R$ is the absorption coefficient of the cladding, that is assumed to remain a constant. Whereas $\mu_c$ is the absorption coefficient of the core, that we can write as 
\begin{equation}\label{eq:mu_c}
    \mu_c(x,y) = \mu_0 + \mu_0 g_\mu (x^2+y^2).
\end{equation}
In this way, we may recover the case of a SIF by putting $g_\mu=0$, and of a GIFs by imposing $\mu_0 = \mu_R$.

The size of the optical fiber is negligible with respect to that of the X-ray beam used in our experiments. Therefore, despite the large divergence angle $\theta$, we can approximate the beam by a plane wave. 
Under the hypothesis that refraction and diffraction are negligible, the intensity of a plane wave $I$ when crossing an absorbing medium follows the Lambert-Beer equation, which reads:
\begin{equation}
    \frac{\partial I}{\partial y} = - \mu (x,y) I (x,y).
    \label{eq:lambert}
\end{equation}
Note that here we are neglecting all edge effects, which occur at the core-cladding and cladding-air interfaces. Eq.(\ref{eq:lambert}) can be easily integrated, thus obtaining:
\begin{equation}
    \ln \Bigg(\frac{I_0(x)}{I(x)}\Bigg) = \int_0^L \mu (x,y) dy,
    \label{eq:lambert_sol}
\end{equation}
where $I_0 (x)$ and $I(x)$ are the intensities of the beam which reaches the optical fiber or the detector, respectively. The integration extreme $L$ is the thickness crossed by the X-ray along the $y$ direction at a distance $x$ from the fiber axis. We highlight that Eq.(\ref{eq:lambert_sol}) allows for clearly understanding the relationship between tomographic and radiographic images. Indeed, by looking at Eq.(\ref{eq:lambert_sol}), it is clear that integrating all of the tomographic reconstructed slices provides the profile of a single radiography, as we stated in Sec.\ref{sec:results}.\ref{sec:radio-tomo}.

By substituting Eqs.(\ref{eq:mu}) and (\ref{eq:mu_c}) in Eq.(\ref{eq:lambert_sol}), after trivial integration, one gets:
\begin{equation}\label{eq:fit}
\begin{array}{ll}
\ln\Bigg(\frac{I_0}{I(x)} \Bigg) & =  2\mu_R \sqrt{R^2-x^2} + 2\mu_0\sqrt{r_c^2-x^2} +\\
    & -2(\mu_R-g_\mu \mu_0 x^2)\sqrt{r_c^2-x^2}\\
     &  +\frac{g_\mu \mu_0}{3} \Bigg[\bigg(\sqrt{R^2-x^2}+\sqrt{r_c^2-x^2}\bigg)^3 + \\
     & \quad\quad -\bigg(\sqrt{R^2-x^2}-\sqrt{r_c^2-x^2}\bigg)^3\Bigg],
\end{array}
\end{equation}
when $x<r_c$. The latter expression can be used for fitting the experimental data, and for extracting information about unknown parameters, such as $\mu_R$, $\mu_0$ and $g_\mu$. 
It is worth to underline that in the presence of plastic coating, the function (\ref{eq:fit}) further complicates due to the presence of an additional unknown absorption coefficient, i.e., that of the jacket. The derivation reported above can be easily generalized to more complex structures, such as that of PCFs. Nevertheless, this goes beyond the goal of our present work, where we limit our analysis to the case of standard optical fibers. 
Here, we fit the experimental data in Fig.\ref{fig:radiography}e and f, and the radiography of sample D (not shown) by means of Eq.(\ref{eq:fit}). This is shown in Fig.\ref{fig:theo}b, from which one can visually appreciate the good agreement between experimental data and the fitting function.

\section*{Calculation of nominal core/cladding refractive index difference}\label{app-delta}
Here, we report the calculation of the nominal values of the parameter $\Delta_n$ for GIFs, which are reported in the main text, and, specifically, in Table \ref{tab:delta}. 
We started by calculating the refractive index of the cladding ($n_{clad}$) by means of the Sellmeier formula, i.e.,
\begin{equation}
    n_{clad} = \sqrt{1+\sum_1^3 \frac{B_i \lambda^2}{\lambda^2-C_i}}.
\end{equation}
We consider the cladding to be made of pure fused silica, thus we used the following parameters: $B_1=0.6961663$, $B2=0.4079426$, $B_3 =0.8974794$, $C1= 0.0684043$, $C2=0.1162414$, and $C3=9.896161$.
Next, we recall the definition of the parameter $\Delta_n$ as  
\begin{equation}
\Delta_n = (n_{0}^2-n_{clad}^2)/2n_{0}^2,
\end{equation}
where $n_{0}$ is the refractive index at the center of the core. The latter is calculated by considering the following formula for the numerical aperture
\begin{equation}
 \mathit{NA} = \sqrt{n_0^2 - n_{clad}^2},   
\end{equation}
where the core and cladding index values are provided by the fiber manufacturer.
Note that $\Delta_n \simeq n_0 - n_{clad}$ as long as $n_0 \simeq n_{clad}$, i.e., $\Delta_n \ll 1$, which is the case of all of the GIF samples considered in this work (cfr. values of $\Delta_n$ in Table \ref{tab:delta}).

\section*{Luminescence of optical fiber defects}\label{app-lum}

\begin{figure}[ht!]
\centering\includegraphics[width=8.7cm]{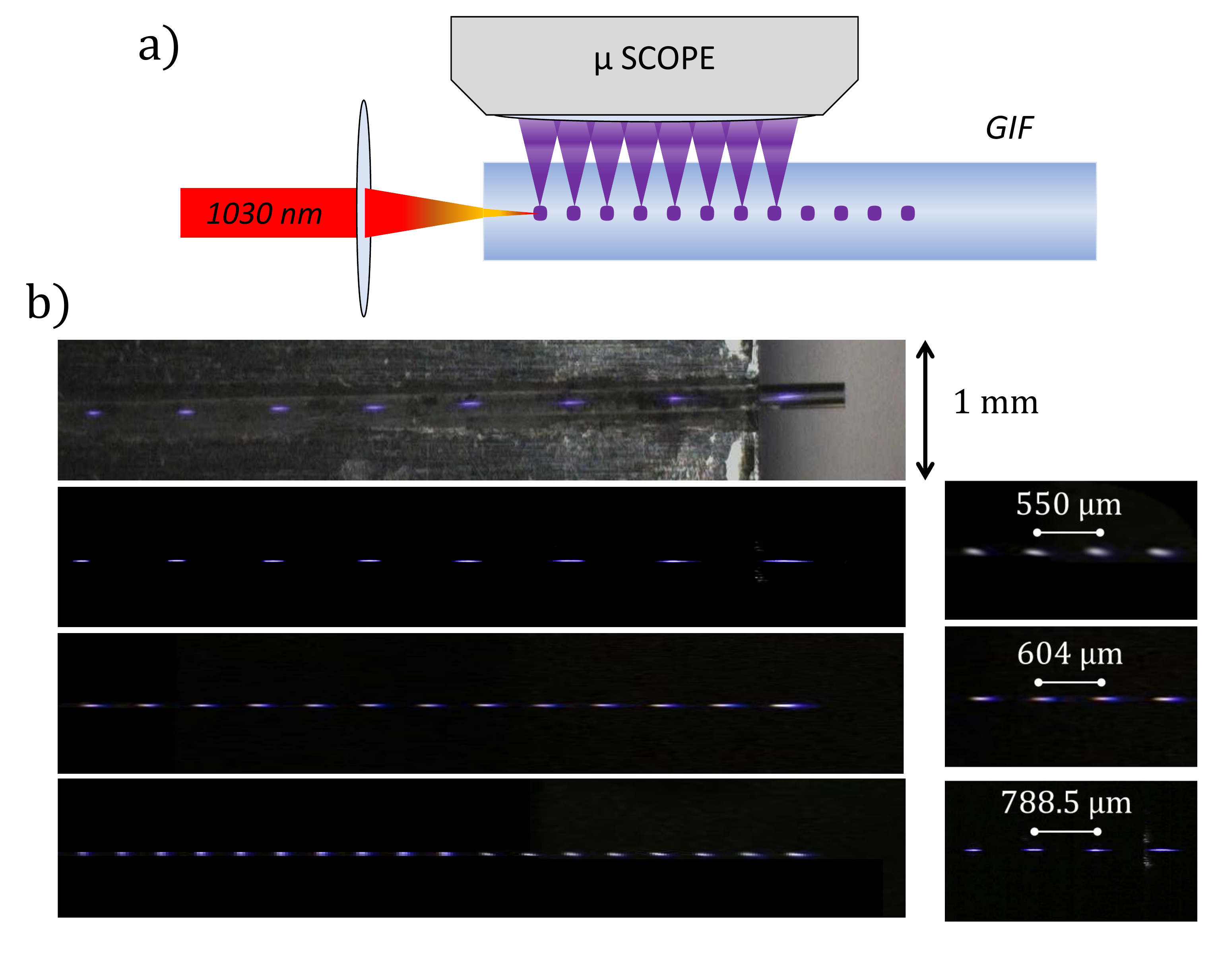}
\caption{a) Sketch of the experimental set-up for the detection of the luminescence of GIF defects. b) Luminescence of samples D, E, and F (from bottom to top) and relative measurement of the periodicity (right panel). The very top figure is the same as that underneath, but with room light switched on.
}
\label{fig:lum-setup}
\end{figure}

Optical fibers exhibit different kinds of defects \cite{girard2019overview} which can be either intrinsic, i.e., associated to pure silica, or due to the presence of doping. Among these, some produce luminescence in the visible spectral region when excited by UV lasers, since their absorption band is in the range of a few electronvolts. Still, the defect luminescence can be excited by infrared lasers via multiphoton absorption processes \cite{mangini2020multiphoton}, which is the case of this work.
Such nonlinear processes require light pulses with peak powers as high as a few Megawatts, which we could reach by means of an ultra-short laser system (Lightconversion PHAROS-SP-HP). This generates femtosecond pulses at 100 kHz repetition rate and $\lambda=1.03$ $\mu$m, with Gaussian beam shape ($M^2$=1.3). 
Laser pulses were injected into the optical fiber core by a convex lens, so that at the fiber input the beam $1/e^2$ of peak intensity is approximately 17 $\mu$m.
The input tip of each fiber, where the luminescence is excited before the laser beam intensity is damped by nonlinear absorption \cite{ferraro2021femtosecond}, is imaged by a digital microscope (Dinolite-AM3113T) as sketched in Fig.\ref{fig:lum-setup}a. In particular, in the special case of a GIF, which has a parabolic index profile, luminescence is emitted at an array of spots, owing to the so-called spatial self-imaging effect which periodically shrinks the beam size, thus enhancing its peak intensity \cite{hansson2020nonlinear}.

This can be appreciated in Fig.\ref{fig:lum-setup}b, where we report the images of samples D, E, and F, which emit a typical violet luminescence. Specifically, the latter is due to the presence of the Ge doping \cite{girard2019overview}.
The periodicity of the luminescence array ($\Gamma$), which is illustrated in Fig.\ref{fig:lum-setup}c, depends on the size of the fiber core, as well as on the difference between the real part of the refractive index at the core center and that of the cladding via the relationship
\begin{equation}
    \Delta_{lum} = \frac{\pi^2 r_c^2}{2 \Gamma^2}.
\end{equation}
Thus, by measuring $\Gamma$, we may compute the values of $\Delta_{lum}$ which are reported in Table \ref{tab:delta} and in Fig.\ref{fig:lum}b.

\end{document}